# TOWARDS A QUANTITATIVE MODELING OF THE SYNTHESIS OF THE PECTATE LYASES, ESSENTIAL VIRULENCE FACTORS IN *Dickeya dadantii*

## Wilfred D. Kepseu[1§], Jacques-Alexandre Sepulchre[1§], Sylvie Reverchon[2] and William Nasser[2]

From Institut Non Linéaire de Nice, UNSA, CNRS UMR 6618, 1361 route des Lucioles, 06560 Valbonne, France[1], Univ Lyon, F-69622 Lyon France; Université Lyon 1 Villeurbanne; INSA-Lyon F-69621 Villeurbanne; CNRS UMR5240 Microbiologie, Adaptation et Pathogénie [2]

§ These authors contributed equally well to this work.

Address correspondence to: Jacques-Alexandre Sepulchre, email: *jacques.alexandre-sepulchre@inln.cnrs.fr*

## ABSTRACT

A dynamic mathematical model has been developed and validated to describe the synthesis of pectate lyases (Pels), the major virulence factors in *Dickeya dadantii*. This work focuses on the simultaneous modeling of the metabolic degradation of pectin in by Pel enzymes and the genetic regulation of *pel* genes by 2-keto-3-deoxygluconate (KDG), a catabolite product of pectin which inactivates KdgR, one of the main repressors of *pel* genes. This modeling scheme takes into account the fact that the system is composed of two time-varying compartments: the extracellular medium, where Pel enzymes cleave pectin into oligomers, and the bacterial cytoplasm where, after internalization, oligomers are converted to KDG. Using the quasi-stationary state approximations, the model consists of some nonlinear differential equations for which most of the parameters could be estimated from the literature or from independent experiments. The few remaining unknown parameters were obtained by fitting the model equations against a set of Pel activity data. Model predictions were verified by measuring the time courses of bacterial growth, Pel production, *pel* mRNA accumulation and pectin consumption under various growth conditions. This work reveals that pectin is almost totally consumed before the burst of Pel production. This paradoxical behaviour can be interpreted as an evolutionary strategy to control the diffusion process so that as soon as a small amount of pectin is detected by the bacteria in its surroundings it anticipates more pectin to come. The model also predicts the possibility of bistable steady states in the presence of constant pectin compounds.

*Dickeya dadantii* (ex *Erwinia chrysanthemi*) is a soft-rotting Gram-negative bacterium that attacks a wide range of plant species, including many crops of economical importance. These bacteria are found on plant surfaces and in soil where they may enter the plant via wound sites or through natural openings. During infection, *D. dadantii* first colonizes the intercellular space (apoplast) where they can remain latent until conditions become favourable for the development of the disease. Soft rot, the visible symptom, is mainly due to the degradation of pectin present in the plant cell wall. *D. dadantii* can utilise pectin as its sole carbon and energy source (Fig. 1). The depolymerization of this polysaccharide, consisting of α-*1,4*-linked galacturonate residues, is accomplished by a variety of pectinases. Among these pectinases, *D. dadantii* endo-pectate lyases (Pels) are known to play the major role in the soft-rot disease caused by this bacterial species *(1,2)*. These enzymes randomly cleave, by β-elimination, the internal glycosidic linkages in pectic polymers, preferentially polygalacturonate (PGA), de-esterified pectin, or low-esterified pectin. They generate a series of unsaturated oligogalacturonates (UGA). The end products of pectin degradation by the extracellular endo-Pels enter the periplasm through the KdgM and KdgN porins and are subsequently internalized into the cytoplasm via the two transporter systems: TogT and TogMNAB *(3-5)*. In this cellular compartment, small oligomers are then degraded into two types of monomers, galacturonate and 5-keto-4-deoxyuronate (DKI), which are both converted to 2-keto-3-deoxygluconate (KDG) (Fig. 1).

Production of the Pels is controlled by a complex regulation system involving multiple transcriptional regulators that respond to various stimuli, such as the presence of pectin compounds, cyclic adenosine monophosphate (cAMP), growth phase, temperature, osmolarity, pH and iron concentration *(6-13)*. Among these conditions, the effect of pectin is predominant *(6)*. The induction of *pel* gene expression by pectic compounds involves a de-repression. The interaction of the repressor KdgR with pectin catabolites, mainly KDG, relieves the binding of this transcriptional regulator to its operators, situated in the vicinity of the promoters of the controlled genes *(14-16)*.

Attention has been paid recently to the theoretical modeling of the onset of virulence in *D. dadantii (17,18)*. From a systems biology perspective, the process of Pel synthesis is interesting as it combines metabolic and genetic regulations. Apart from a few exceptions *(19-21)*, the simultaneous modeling of metabolic and genetic interactions in the same network is seldom performed because it involves variables that evolve on very different time-scales. In bacteria, the typical response time of a metabolic pathway can be less than a few seconds, whereas the response time for protein synthesis is of the order of several minutes at least. On the other hand, feedbacks often exist between both networks because the metabolic reactions rely on enzymes encoded by genes whose expression can be regulated by products of the metabolic pathways. For example, the production of the Pel enzymes enables the catabolism of pectin, leading to the production of KDG which, in turn, sequesters one of the main repressors of *pel* genes. The first study concerning Pel synthesis was carried out by Sepulchre *et al*. *(17)*. This work made it possible to put together all the information from various experiments (each of the experiments concerning, in general, only one regulator) in the form of a coherent transcriptional regulatory network. This offers the possibility of assigning a set of mathematical equations describing the dynamics of the full network *(17)*. Using this qualitative mathematical modeling, new insights have been gained regarding the infection process. In particular, a hierarchy of the relative importance of the numerous regulators experimentally studied has been deduced from the theoretical analysis of the network. This qualitative analysis highlighted that the minimal, but essential, subnetwork underlying Pels synthesis is the feedback loop involving the repressor KdgR and the metabolic aspect of the network. In another study *(18)*, considering only the identified subnetwork, the authors showed that, in principle, this KdgR loop was enough to predict various possible phenomena, such as bistability, excitability and autonomous oscillations. It should be noted that the precise dynamical regime here is crucially dependent on the actual values of the parameters *(18)*. This subnetwork was already thought to be essential by the biologists but the modeling served to confirm its importance *(6,14,16)*.

The problem with the qualitative models *(17,18)* is the lack of knowledge concerning the parameters. This shortcoming hampers further exchanges between the theory and the experiments, in particular for predicting the phenomena that could occur during virulence. So, our aim in this work is to achieve a more quantitative mathematical model of Pels synthesis which could account for most of the current experimental data obtained with variable concentrations of PGA, thus validating or invalidating our current understanding that the key mechanism of the induction of Pels synthesis is associated with the KdgR positive feedback loop. Until recently, no precise information about substrate degradation and the induction of Pels synthesis was available. Unexpectedly, we observed from experiments that PGA is almost totally consumed, whatever the concentration used, prior to the induction of Pels synthesis. It was also observed that, when changing the initial PGA concentration, the Pels production profile is relatively constant but the maximal Pels concentration values which can be reached depend on this initial amount. Another question arising from the previous modeling approach concerns experiments on the dynamics of transcripts of the *pel*. Our recent experimental results reflect that the dynamics of transcripts are very fast compared with that of the proteins. So, the decay of Pels induced by its degradation rate, after reaching their maximum level, is mainly due to the absence of mRNA production.

The triggering of the synthesis of Pels by the KdgR loop can be qualitatively understood as follows. When a small amount of Pels is produced it induces the degradation of pectin and its transformation into KDG (Fig. 1 and Fig. 2). The KDG subsequently causes the inactivation of KdgR, leading to a complete removal of the action of KdgR on *pel* genes. To develop a quantitative model, the following aspects have now been taken into account and added to the previously simplified subnetwork *(18)*: the degradation of KdgR, despite being complexed with KDG-KdgR, and the compartmentalisation of the medium, i.e inside and outside the bacteria. In other words, we now take into account the export of Pel proteins by the Out secretion system and the import process of oligogalacturonate (UGA) (Fig. 1). Fig. 2 represents the transcriptional regulatory subnetwork of *pel* using the graphical conventions proposed in *(21)*.

## EXPERIMENTAL PROCEDURES

*Bacterial Strain and Culture Conditions* - *D. dadantii* 3937 *(23)* was used for all the experiments described. Cultures were grown at 30°C in M63 minimal salt medium *(24)* supplemented with a carbon source: glucose at 0.2% (w/v) and various concentrations of polygalacturonate (PGA), from 0.05 to 0.4% (w/v). Liquid cultures were grown in a shaking incubator (220 r.p.m.).

*Enzyme Assays* - Assay of pectate lyase was performed on supernatants of the bacteria cultures or on toluenized cell extracts. Pectate lyase activity was determined by monitoring spectro-photometrically the degradation of PGA to unsaturated products that absorb at 230 nm *(25)*. The molar extinction coefficient of unsaturated oligogalacturonates used was 5200 (i.e. 1 mole/l gives an absorbance at 230 nm of 5200) *(25)*. Pel activity is expressed as µmoles of unsaturated products liberated per min and per ml of enzymatic extract. The standard assay mixture consisted of 100 mM Tris-HCl (pH=8.5), 0.1 mM $CaCl_2$, and 0.5 g of PGA per liter in a total volume of 1 ml. Bacterial concentration was estimated by measuring turbidity at 600 nm, given that an absorbance at 600 nm ($A_{600}$) of 1.0 corresponds to $10^9$ bacteria per ml and to 0.47 mg of bacteria (dry weight) per ml *(26)*. In some cases, particularly to quantify the enzyme activity in a certain growth medium condition, the Pel assays were performed in M63 minimal medium (pH=7.0) supplemented with $CaCl_2$ and PGA, as above.

*Monitoring of PGA Degradation During Growth of the Bacteria* – An indirect strategy was used to follow PGA degradation. Briefly, bacteria were grown in liquid M63 medium containing glucose (0.2% w/v) and PGA (0.01 to 0.4% w/v) and 10 ml samples were taken every hour. Cells were harvested by a centrifugation of 5 min at 6 000 x g, then 6 ml of the supernatant was taken and submitted to an incubation of 20 min at 95°C to inactivate the Pels contained in the medium. The resulting samples were used as PGA sources for the preparation of Pel assay buffers, as described above. Aliquots of the media taken before inoculation with the bacteria, and treated as described above, were used as a reference.

*Protein Cellular Concentration and Degradation Rates Measured by Antibiotic Chase* – Overnight cultures were inoculated into fresh M63 minimal medium containing glucose (0.2 % w/v) and supplemented, or not, with PGA (0.2 % w/v). At the late-exponential phase ($A_{600}$=0.8 for M63+ glucose and $A_{600}$=1 for M63+glucose+PGA), chloramphenicol was added to a final concentration of 200 µg/ml from a freshly prepared stock solution. Samples (1ml) were removed between 0 and 48h, centrifuged, resuspended in an adequate volume of Laemmli sample buffer and boiled for 2-3 min. Samples were separated by SDS-PAGE (12% polyacrylamide) and transferred for 20 min onto a nitrocellulose membrane, using a semi-dry blotter. Western blots of the separated proteins were incubated with a polyclonal anti-KdgR or anti-Pels antibody as the primary antibody, and an anti-rabbit peroxidase-conjugated antibody (SIGMA) as the secondary antibody. The signals obtained in Western experiments were detected by autoradiography on Amersham MP film and quantified using ImageMaster TotalLab version 2.01 software (Amersham Biosciences). The protein concentration in bacteria cells was determined by densitometric analysis and then a comparison of the signals was obtained with a serial dilution of the purified protein.

*RNA Isolation and Quantitative Reverse Transcription Polymerase chain Reaction (RT-PCR) Analysis* - Total RNA was extracted from *D. dadantii* by the Qiagen Rneasy Mini kit procedure (Qiagen). RNA was quantified using a Nanodrop spectrophotometer and then checked on a 1%-agarose gel containing 0.5 µg/ml ethidium bromide. The cDNAs were synthesized with 1 µg of RNA using the SuperScriptTM first-strand synthesis system for RT-PCR (Invitrogen), in the presence of 50 ng random hexamers/µg of RNA. The reaction mixture was incubated at 25°C (10 min), 42°C (50 min) and 70°C (15 min). 2.5 x $10^5$ copies of Gene Amplifier pAW $10^9$ RNA (Applied Biosystems) were added to the reverse transcription reaction and used as a control for retro-transcription efficiency. 0.1 to 1 µl of the reaction mixture obtained was used for q-PCR reactions in a volume of 10 µl, using the LightCyclerR faststart DNA masterplus SYBR Green I kit from Roche (Roche Applied Science).

The real-time PCR reaction was performed in a Roche LightCycler 480. Reactions were performed at 95°C for 10 min and 35 cycles of 95°C for 15 s, 55°C for 15 s and 72°C for 20 s. Target gene expression is defined by the method described by Pfaffl *(27)*. The *rsmA (11)* and *rpoA* genes were selected as the reference genes for real-time RT-PCR for an accurate normalisation.

*Model formulation* – We examined most of the biochemical reactions occurring between molecules of the network, represented in Fig. 2. It is important to note that, in this system, the enzymatic action of Pels takes place in the extracellular medium, whereas the transcriptional regulation by KdgR and its inactivation by KDG occur inside the bacterium (Fig. 1 and 2). To set up the equations the CellDesigner environment *(28)* was used to represent the chemical reactions in the form of an interaction graph (Fig. 3), from which ordinary differential equations (ODEs) describing the evolution of the variables are automatically generated using standard mass action laws. The set of equations obtained is reported in the Supplemental Material. Next, the dimension of the system is reduced by considering that the fast variables entering into the set of equations are at their equilibrium points. Although this so-called quasi-steady state approximation (QSSA) is standard in modeling biochemical kinetics, the identification of the slow variables in molecular networks can be difficult in practice because most of the molecules participate, in fact, in both slow and fast dynamics. In our case, however, this separation is relatively straightforward to operate by considering slow variables as representing the total concentrations of proteins, including the free and all the complexed forms of those proteins. This procedure enables us to reduce the kinetic equations to a set of just a few ODEs, describing the dynamics of the concentrations of the (total) proteins Pels, KdgR and of the carbohydrates PGA, UGA and KDG. These variables are denoted, respectively, by ($x$, $y$, $s$, $z$, $w$) in the equations below. Model simplification is detailed in the supplemental material. As shown on Fig.1, the product of PGA degradation consists of unsaturated oligogalacturonates (UGA) which are obtained in the extracellular medium, imported into the cells *(3-5)* and then transformed into KDG.

PGA is a polysaccharide polymer whose enzymatic fragmentation into oligomers is a multistep process, involving in principle the degradation of polymers with any intermediate sizes between the longest polymers (size $L$) and monomers. To model this biochemical complexity, we have considered two types of approaches: on one hand, we have described a complete set of enzymatic degradation reactions, starting with a distribution of polymer sizes with a mean length of 210 and maximal length $L=280$, which corresponds to the PGA used in the experiments. On the other hand, in order to compare the full polymer degradation process with a minimal model, we have also considered the case of only one enzymatic reaction, i.e. starting from dimers ($L=2$) that are cleaved into monomers. As discussed below, experimental data can be fitted equally well with both types of modelings. So we will only write the equations of the simplest model in the following. Nevertheless the equations of the more detailed model are presented in the supplemental material and the parameter values obtained with both models are given for comparison in Table 2. The model equations can be summarized as follows:

$$\frac{dx}{dt} = \beta \left( \frac{\rho}{1-\rho} \right) \frac{K_{d6}}{K_{d6} + y_d} - \alpha_1 \frac{K_m}{K_m + s} x \qquad (1)$$

$$\frac{dy}{dt} = \beta_2 - \left( \alpha_2 + \frac{\dot{\rho}}{\rho} \right) y \qquad (2)$$

$$\frac{ds}{dt} = -\varepsilon k_{cat} x \frac{s}{K_m + s} \qquad (3)$$

$$\frac{dz}{dt} = 2\varepsilon k_{cat} x \frac{s}{K_m + s} - \gamma \rho z \qquad (4)$$

$$\frac{dw}{dt} = \gamma(1-\rho)z - \left( k_s + \frac{\dot{\rho}}{\rho} \right) w \qquad (6)$$

$$\frac{d\rho}{dt} = \sigma \rho (1 - \frac{\rho}{\rho_s}) \qquad (7)$$

with $y = 2 y_d \left[ 1 + \frac{w^2}{K_{d3}^2} \right]$

In these equations the variable $x$ represents the concentration of Pel enzymes. However, for comparing the results of numerical simulations with the experimental data it is custom to show instead of $x$ the quantity $k_{cat}x$, called the *enzymatic activity* that is expressed in units of mmol.min$^{-1}$. Thus, this is the variable that will be drawn on the Figures presented below. The variable $y_d$ represents the concentration of free dimers KdgR$_2$. The variable $s$ stands for the concentration of the substrate PGA, whereas $z$ is the end product of degradation of *s*, corresponding here to the concentration of monomers that will be subsequently transformed into KDG (*w*). In reality the end product of degradation of PGA by the Pels are dimers (unsaturated digalacturonates) that are transported into the cytoplasm where other enzymes than Pels perform the monomerization reactions (Fig.1). In order to simplify these steps, we model the complete depolymerization of PGA in the external medium. Then the importation of monomers into the cytoplasm is described at once by the negative term indicating a consumption in eq.(4). Let us notice that this term must be proportional to the number of cells in the medium. The variable *w* represents the KDG concentration, whose decay is mainly due to the phosphorylation of KDG. Finally the variable $\rho$ is the bacterial volume fraction whose growth is governed by a logistic equation *(29)*, in which $\sigma$ is the bacterial growth rate and $\rho_s$ is the maximum bacterial volume fraction. The notation $\dot{\rho}$ in the equation (2) denotes the time-derivative of $\rho$. Let us notice that the bacteria growth depends on the presence of two substrates, PGA and glucose. In this study, the glucose is maintained constant in each experiments and the effect of varying PGA substrate on the Pel dynamics is studied. The dependence of the parameters of the bacteria growth (as shown on Suppl. Material) on the initial amount of PGA substrate ($s_0$) is obtained to be linear and given by:

$$\sigma = \sigma_0 + a s_0 \qquad (8)$$

$$\rho_s = \rho_0 + \rho_{s_0} + Y_p s_0 \qquad (9)$$

The parameters $\sigma_0$ and $\rho_{s_0}$ depend on the quantity of the second substrate (glucose) which is kept fixed in this study. Therefore, at the beginning of the experiment, knowing the initial bacterial volume fraction ($\rho_0$) and the amount of substrate introduced initially ($s_0$) can allow to solve the full set of eqs.(1)-(7). Indeed, each parameters of the set of equations (1)-(7) is defined in Table 1 and Table 2. All computations and visualisations are performed in MATLAB R2007b.

## RESULTS

*Parameter identification* - In order to analyze our mathematical model, we evaluated the numerical values of the parameters of our model. The parameters are either estimated using data from the literature or fitted using experimental data. In the following section, we give details about the specific experiments that were required in order to determine some of the parameter values in the set of equations (1)-(5).

Fig. 4 shows, on the same graph, three types of data measured from the same experiment, which was performed in conditions of Pel induction with initial PGA (0.01% w/v), and glucose (0.2% w/v) in minimal salt medium M63. The graph shows the experimental data of bacterial growth ($A_{600}$), Pel enzymatic activity and the transcripts of *pelD*, which is a good representative of the family of *pel* genes in the induction conditions studied. On the same graph a continuous curve fits the Pel activity by using the simplest differential equation model relating the transcripts and the protein data. This model, which can be formulated as follows,

$$\frac{dx_1}{dt} = a\,m(t) - \alpha_1 x_1,$$

shows that the protein concentration evolves with a production term that is proportional to the transcript quantity $m(t)$, and that its degradation is proportional to the protein concentration. Here the transcripts are measured in arbitrary units, so the factor *a* is also arbitrary. The relevant information which can be extracted by parameter optimization of this elementary model is the degradation parameter $\alpha_1$. The value obtained for this parameter tells us that the half-life of Pels is 15 hours, quantifying the fact that the Pels proteins are known to be quite stable.

In Fig. 4, the bacterial growth, measured by the variations of the $A_{600}$ value, is fitted by means of the eqs.(7)-(9). The inflexion point of the fitting curve defines the transition time between the exponential and the stationary phase, and is found to be $t_m$= 7.34 *h*. On the other hand, the point of maximum production of the transcripts ($t$=8.5 *h*) occurs later than $t_m$. This proves that induction of the *pel* gene expression takes place at the beginning of the stationary phase.

Another parameter appearing in the model equations is the Michaelis Menten (MM) constant, $K_m$. The weight-characterized MM constant, which we denote in the following by $K_M = K_m.P$, where *P* is the molecular weight of the galacturonate (see Table 1), had previously been measured for the five major Pel isoenzymes (*A-E*) in standard optimized buffers *(30)*. Therefore a value of $K_M$=0.2 g/l, corresponding to the mean value of $K_M$ amongst the Pel enzymes (*A-E*), was initially used for the modeling. This value was found to be inappropriate because the actual affinity of the Pel enzymes for their substrate (PGA) is significantly lower in the culture medium (M63 minimal medium, pH =7.0) than in the standard buffer assays (Tris-HCl pH= 8.5) used in *(30)*. Hence new enzymatic assays were performed using the M63 culture medium as buffer (see experimental procedures). These assays corresponded well with the classical MM law provided the PGA concentration was not too high. A value of $K_M=1.2$ g/l was deduced from the half maximum enzymatic activity, as seen on the graph shown in Fig. 5. Notice on this graph that the MM law fails to describe the experimental data when the PGA concentration exceeds 5 g/l, which indicates that the enzymes become inhibited when the substrate concentration is too high. Table 1 gives the list of parameters deduced from independent experiments and the literature.

Some other parameters could not be experimentally determined and these are considered as being adjustable. The adjustable parameters are determined by minimizing a quadratic error function that represents the distance between the normalized experimental and model data sets (see details in Suppl. Material). This enables us to obtain a set of optimal parameters. Table 2 summarizes the list of parameters which have been determined from the optimal fitting by numerical simulations.

*Total consumption of* PGA - In its virulent phase, one of the main activities of *D. dadantii* is to degrade the pectin which is found in the plant cell walls. However, the quantity of pectin available to a bacterium *in planta* is difficult to estimate. So, in order to quantify the amount of substrate degraded by the bacteria, measurements were performed on samples from a flask containing the M63 medium which mimics the medium encountered by *D. dadantii* in the early steps of infection in the apoplasts. Moreover, in this artificial medium, the pectin is replaced by polygalacturonate (PGA).

Several experiments were performed to observe the effects of variation of the PGA concentration on the dynamics of the *pel* genes expression and on the production of the enzymes. The model was used in order to fit experiments to levels of PGA consumption and Pels synthesis. We considered two experiments carried out in M63 medium, with initial glucose (0.2% w/v) and various initial concentrations of

PGA (0.4% and 0.2% w/v, respectively). The experimental results showing PGA consumption and Pels synthesis are shown in Fig. 6. Considering the initial conditions determined analytically (see suppl. material) and corresponding to the equilibrium of the chemical network (except for PGA, the level of which is increased at the beginning of the experiment), we could numerically solve the simplified equations describing our network. An initial experiment (shown on Fig. 6A) was used to determine the optimal set of unknown parameters, minimizing the quadratic average error between the experimental data obtained and our numerical results (see Suppl. mat). Then, the acquired parameters (see Table 2) were used in order to numerically fit all the other experiments. As seen in Fig. 6A, the initial amount of PGA does not much influence the profile of PGA consumption or Pels synthesis. However, the maximum activity of Pels depends on this initial amount. As can be seen in Fig. 6B, the model can reproduce the experimental results with different amounts of PGA. After about 8h, the amount of free PGA is nearly completely consumed before the synthesis of Pels begins.

*New addition of PGA* – Next, we consider a series of experiments in which additional amounts of PGA are added 24h after inoculation of the cultures. At this time the initial PGA is completely consumed and the Pels concentration has already passed its maximum level. Fig. 7 shows the dynamics of three different experiments in which PGA (0.01%, 0.025% and 0.05% w/v) is added. In this case, it is apparent that the maths model is also well adapted for fitting with the corresponding experiments. The addition of PGA after 24h results in a slight revival of bacterial growth (not shown) and in the resumption of Pel activity. As observed in the inset in Fig. 7, the intensity of these two phenomena, in particular the induction of Pel activity is proportional to the concentration of added PGA.

*Study of parameter sensitivity* – Biochemical parameters obtained from the literature or from independent experiments, as those listed in Table 1, are usually presented as error free point estimates. In fact, these values are often obtained from indirect measurements, so that they could suffer from some inaccuracies. Moreover these values could sometimes be fluctuating from one experiment to another. In order to test the sensitivity of the model predictions with respect to fluctuations of these parameters, further simulations have been performed, where the relative variability on the established parameters could reach 20% around the value provided in Table 1. The value taken by any parameter $p$ in Table 1 is randomly chosen in an interval ($p_{min}, p_{max}$) defined around its mean value given in this Table. Fig. 8 shows the obtained extreme curves between which all fits are confined when the considered parameters are changed with a relative variation of 10%. Let us remark that the upper and lower limits of the "gap curves" correspond respectively to either the $p_{min}$ or the $p_{max}$ values of varied parameters. Furthermore, for any random choice of parameters within the 20% variation intervals, it is possible to obtain a choice of parameters of Table 2 (by applying the same optimization procedure) which allows to achieve a fit of experiments data very close to the blue curve of Fig. 8.

## DISCUSSION

Our quantitative modeling captures the main features of the time evolution of Pels synthesis by *D. dadantii* after induction by PGA. An original feature of our modeling scheme is that the system is composed of two dynamic compartments. The first compartment corresponds to the total bacterial volume and the second to the supernatant in the culture medium. Both parts of the system are equally important since the pectate lyases mainly carry out their enzymatic actions on their substrate in the supernatant, whereas the product of their degradation is imported into the bacteria. Because the external volume is much greater than the bacterial volume, the export and import processes are followed, respectively, by a significant dilution or concentration of the chemical species concerned. Moreover, these changes in concentration are time-dependent, since the bacterial volume is increasing with time, evolving via the two classical phases observed for a growth-limited population, namely the exponential and the stationary phases. The former is characterized by σ, the exponential growth rate of the population, and it is observed that this parameter plays a major role in the onset of virulence factors. Indeed a prediction of our model obtained by numerical simulations is that the time at which the maximal concentration of Pels shows up is strictly proportional to the generation time of bacteria, as illustrated on Fig. 9. Incidently, this Figure also shows that the time of maximisation of KDG shares the same property. As a matter of fact there is a substantial time-delay between the moment of inoculation of bacteria and the time of maximisation of KDG concentration. One can interpret this delay as resulting from the fact that the feedback loop between the degradation of PGA allowing the production of KDG and the subsequent derepression of *pel* genes is strongly driven by the growth of the bacteria. In particular, the importation of UGA is described in eq.(4) by a consumption term which is proportional to the biomass, allowing an accumulation of UGA in the supernatant during the early exponential phase. Another global regulation of the gene expression by the growth phase is due to the dilution occurring in the cytoplasm during the exponential phase, since the bacterial volume is doubling at each generation. For example this dilution effect is responsible for a temporary diminution of the concentration of KdgR (see Fig.8) and this appears to significantly contribute to the derepression of *pel* genes. In absence of this effect less than half of the Pel production would occur.

As mentionned earlier we considered two ways of modeling the PGA degradation. The first option (details in the suppl. Material) was to take into account all the intermediate steps between the longer polymer size (e.g. L=280) and the smallest oligomers of unsaturated galacturonates. The second option was to deal with a minimal model with only one enzymatic reaction corresponding to the degradations of dimmers into monomers (eqs.(3)-(4)). The two models can fit the data with a similar accuracy, by using the same set of parameters except for the enzymatic reaction rate $\varepsilon k_{cat}$. If the same parameter $\varepsilon$=1 is used for both values of L (L=2 and L=280), an advance of about 1,5 hours is seen on the consumption of PGA, as compared with the same initial condition but with long polymers. In fact in the case L=2 this time advance can be compensated by reducing the enzymatic reaction rate by a factor 3. (i.e. $\varepsilon$=1/3). Therefore for the presentation and the discussion of the model we preferred adopt the case L=2, with an apparent reduced $\varepsilon k_{cat}$, but with a gain in simplicity and in speed of numerical simulations. A compromise would be to simulate a system with moderate polymer size but still larger than 2 (e.g. L=10). In this case there would be practically no time difference in the degradation time of PGA, and still the system would keep a low number of variables.

Our model is robust in the sense that a unique set of parameters was determined using the data shown in Fig. 6*A*. Then, by using the same parameters, the numerics can predict the behaviour of the system under different initial conditions, such as in the cases represented in Fig. 6*B* and Fig. 7. For example, in past studies of induction experiments of *pel* genes, the consumption of the substrate was not monitored. With our current modeling our previous experimental data can be complemented retrospectively with additional variables since the numerical simulations give access to non-measured data, such as the consumption of PGA. It should be noted that the simulations of substrate consumption slightly overestimate the degradation of PGA as time evolves. For example, in Fig. 6*A* the fall of the continuous curve fitting the measured PGA is steeper than the experimental data. This observation is systematic for other data (not shown) and we conclude that there is an unexplored phenomenon, not captured by the present model, that causes a decrease in the enzymatic activity of Pels when these proteins accumulate.

The measure of the kinetics of PGA in time has revealed another striking feature, which is visible in Figs. 6. To explain this point we need to consider the intersection of the two curves (PGA and Pel activity in time), which occurs at around 8.5 hours after the inoculation. At this time, 95% of the initial amount of PGA added has already been consumed, whereas the production of Pels has only reached 5% of the maximal level and it continues to increase. Thus, at first sight, the behaviour of the bacteria does not seem to be optimal, because a major proportion of the Pels production is wasted and not used to degrade PGA. On the contrary, one would expect that in bacteria most enzymatic production would cease as soon as the substrate is absent (e.g. in the classical example of the *lac* operon, the production of β-galactosidase is swiftly repressed when lactose is absent). However in this study a massive production of Pels follows the extinction of the substrate. So, the synthesis of Pels appears as an overproduction compared with the needs of the bacteria population, especially because once they are produced the Pels, the latter remain in the medium for hours (Pels being very stable proteins).

To explain this paradox it is important to remember that Pels are virulence factors, and that our measurements were not made in the natural environment of the pectinolytic bacteria. Indeed, these bacteria live in a natural environment that is not closed, so the massive production of Pels may be an evolutionary strategy: the response of the bacteria could be exaggerated because the apparently overproduced Pels are meant to diffuse over a large distance, relative to the emitting bacteria. Thus the overproduction of Pels could be an "environment-learned" characteristic of the bacteria. In a sense, this is a form of anticipatory behaviour on the part of the bacteria, which is similar to other forms of bacterial anticipations recently observed in *Escherichia coli (31)*. For example, the team of S. Tavazoie has found 'anticipative' behaviours in culture media, where the anaerobic mode is switched on by raising the temperature, even if the oxygen level is maximal *(31)*. Their interpretation is that when these bacteria enter the mouth, the temperature is higher and this is a signal that the oxygen level will soon fall in the digestive tract. In the same way, it seems that the Pels regulatory system has evolved similarly so that the pectin compounds are detected as being a strong signal for the presence of the *D. dadantii* host. So, as soon as a small amount of pectin (or PGA) is detected by the bacteria in its surroundings, it anticipates the arrival of more pectin. A possible test of this hypothesis would be to check whether successive bacterial generations, cultured in the presence of PGA, gradually lose their capacity to synthesize a massive amount of Pels in response to the subsequent addition of pectin to the growth medium.

Another trait that is highlighted by the modeling is the central role of KDG in Pels dynamics. Incidently, the model is able to estimate quantitatively the actual kinetics of KDG (in g/l) that would be present in the medium. Such a prediction may be useful since this variable has not, so far, been experimentally quantified. The bacteria can use KDG in two different ways. A part of these molecules, produced from the degradation of PGA, can bind to KdgR in order to derepress the expression of its target genes, such as the *pel* genes. The other part of the KDG molecules is catabolized and serves as a source of carbon and free energy for the metabolism of the bacteria (Fig. 1). As regards its first function, the binding of KDG to a KdgR dimer can occur either directly on the free protein or on the KdgR already bound to the operators of the *pel* genes. In elaborating the model we first took this second possibility into account but it lead to more complex equations with more unknown parameters, without improving the data fitting. Therefore we only described the sequestration of the free KdgR by KDG to keep the modeling as simple as possible. Regarding its metabolic function, KDG enters into a catabolic pathway by its phosphorylation, via the enzymatic action of KDG kinase (KdgK) *(32)*. In the eq.(6) this transformation of KDG is modeled as a first-order reaction with a constant phosphorylation rate $k_5$. Let us note that this approximation could be improved in a further model because the kinase KdgK is also regulated by KdgR, so the rate of phosphorylation of KDG could vary in time. In the *kdgK* mutant, however, the KDG is no longer catabolized and the excess can sequestrate more KdgR repressors by binding to them. In our mathematical description, the *kdgK* mutant can be modeled by drastically reducing the phosphorylation rate of the KDG. In this case, the model equations could be useful in predicting the behaviour of such a mutant. Fig. 10 shows an example of a numerical result where $k_5$ has been greatly reduced, compared with the value obtained in the wild type. In this situation the synthesis of Pels starts earlier and reaches a higher maximum production than that obtained with the wild type, due to the accumulation of KDG. Let us remark that in this case we must consider a numerical simulation with the growth parameters (eqs.(8)-(9)) corresponding to zero substrate, in order to take into account the fact that blocking PGA catabolism results in bacterial growth conditions with glucose only.

Finally, our modeling can help to predict how the system would behave in a different experimental setting. For example, it is possible to describe the stationary states of the system in a situation where the PGA concentration and the bacterial volume fraction would no longer be variable but considered as control parameters. In fact, the situation of a permanent intake of pectin by the bacteria is somehow met by these organisms in the plant, although this intake is fluctuating. In the case of PGA and bacterial volume being fixed at constant levels, the production of Pels reaches a stationary value that can be estimated from the model by computing the steady states of the system equations. Fig. 11 shows the behaviour of the steady states as a function of the constant PGA, displaying the phenomenon of bistability. In this case there exists a range of PGA parameters where there are two possible stationary states of Pel activity for the same value of PGA. This has an interesting biological implication because the property of bistability is an elementary form of memory *(33)*. Indeed, in the context of our model, this property means that if the bacteria have been in contact with a high amount of substrate and if the available PGA is then reduced the bacteria continues to produce a large amount of Pels, compared with a bacteria that has not encountered PGA before. From the point of view of bacterial virulence, the bistability property would be advantageous for the expression of pathogenicity in *D. dadantii*. Indeed, in this case, assuming that the bacteria hosted in a plant have produced enough KDG by pectin degradation to switch to a massive production of Pels, this property of an underlying memory would guarantee that if, for some reason, after the transition to virulence there is a fall in the concentration of the available pectin, then the level of virulence enzyme production by the bacteria can remain high.

## CONCLUSION

In summary, we have achieved a first quantitative model able to predict the production of Pels in a population of bacteria growing in presence of various amounts of initial PGA. The model is minimal in the sense that it takes into account only two essential ingredients of the regulation of *pel* genes, namely the positive feedback created by the derepression of KdgR by the *pel* inducer KDG and the driving of the kinetic chemical equations with the growth of the bacterial volume. The model is robust because it allows to fit new experimental data with a set of fixed parameters, allowing nevertheless the latter to be varied in a range of 20% relative variations. The model is useful, as it allows to predict the behavior of the Pel regulatory system in situations that are not yet been experimentally investigated. Although the fitting of experimental data is reasonably good, the matching between experimental measurements and the numerics remains moderate. It is true that several aspects of the regulation of the *pel* genes have not been taken into account in our modelisation. For example the fluctuations of the known repressors

PecT and PecS acting on the *pel* promoter modulates the action of the main repressor KdgR. The role of the protein Fis has also been neglected, although this regulator represses the *pel* genes during the exponential phase *(10,11)*. Also, other positive regulators are known to directly impact the production of Pels, like the histone-like protein H-NS and the global regulator CRP. Furthermore the expression of *pel* genes is also affected by physical factors, like the supercoiling state of the bacterial genome *(34)*. This latter physical property is regulated by proteins such that Fis and H-NS, but also by the specific proteins Gyr and TopA. Eventhough in our model a part of the actions of these factors is implicitly taken into account by coupling the biochemical kinetics with the cell growth dynamics, it remains that by choosing not to describe the dynamics of all the regulatory factors we make the implicit assumption that their variations can be discarded or that their effects can be incorporated in the (effective) parameters. But these suppositions are clearly approximations. Therefore our simplifying assumptions are responsible for a part of the discrepancy found between the numerics and the data. The other part should be sought in the intrinsic stochasticity of biological variables.

Our study takes place in a new vein of systems biology research, combining genetic regulations with metabolic processes in the same model. This first attempt will be further developed in a future work by incorporating the role of variation of initial glucose in the model, whereas the latter substrate was constant along the present work. Another perspective that will be developed in a near future will be to study the regulatory dynamics of Pels synthesis with the help of a chemostat.

*Acknowledgments – This work was supported by the ANR program "Regupath" (N°ANR-07-BLAN-0212) which is financed by the French Ministry of Research. In particular W.K. has a post-doctoral contract supported by this funding. Axel Cournac is thanked for stimulating and fruitful discussions, especially concerning the possible bistability in the system.*

# REFERENCES

1. Collmer, A., and Keen, N.T., (1986) *Annu. Rev. Phytopathol.* **24**, 383-409

2. Hugouvieux-Cotte-Pattat N., Condemine, G., Nasser, W., and Reverchon, S. (1996) *Annu. Rev. Microbiol.* **50**, 213-257

3. Blot, N., Berrier, C., Hugouvieux-Cotte-Pattat, N., Ghazi, A., and Condemine, G. (2002) *J. Biol. Chem.* **277**, 7936-7944

4. Condemine, G., and Ghazi, A. (2007) *J. Bacteriol.* **189**, 5955-5962

5. Hugouvieux-Cotte-Pattat, N., and Reverchon, S. (2001) *Mol. Microbiol.* **41**, 1125-1132

6. Reverchon, S., Nasser, W., and Robert-Baudouy, J. (1991) *Mol. Microbiol.* **5**, 2203-2216

7. Reverchon, S., Expert, D., Robert-Baudouy, J., and Nasser, W. (1997) *J. Bacteriol.* **179**, 3500-3508

8. Nasser, W., Robert-Baudouy, J., and Reverchon, S. (1997) *Mol. Microbiol.* **26**, 1071-1082

9. Reverchon, S., Bouillant, M.L., Salmond, G., and Nasser, W. (1998) *Mol. Microbiol.* **29**, 1407-1418

10. Lautier, T. and Nasser, W. (2007) *Mol. Microbiol.* **66**, 1474-1490

11. Lautier, T., Blot, N., Muskhelishvili, G., and Nasser, W. (2007) *Mol. Microbiol.* **66**, 1491-1505

12. Nasser, W., and Reverchon, S. (2002) *Mol. Microbiol.* **43**, 733-748

13. Franza T, Michaud-Soret I, Piquerel P, and Expert D. (2002) *Mol. Plant Microbe Interact.* **15**, 1181-1191

14. Nasser W, Condemine G, Plantier R, Anker D, and Robert-Baudouy J. (1991) *FEMS Microbiol. Lett.* **65**, 73-78

15. Nasser, W., Reverchon, S., and Robert-Baudouy, J. (1992) *Mol. Microbiol.* **6**, 257-265.

16. Nasser, W., Reverchon, S., Condemine, G., and Robert-Baudouy, J. (1994) *J. Mol. Biol.* **236**, 427-440

17. Sepulchre, J.A., Reverchon, S., and Nasser, W. (2007) *J. Theor. Biol.* **244**, 239–257

18. Kepseu, W.D., Woafo, P. and Sepulchre, J.A., (2010) *J. Biol. Syst.* **18** (1), 1-31

19. Ropers, D., de Jong, H., Page, M., Schneider, D., and Geiselmann, J. (2006) *Biosystems* **84**, 124–152

20. Ropers, D., Baldazzi, V., and de Jong, H. (2010) *ACM/IEEE Trans. on Comput. Biol. and Bioinf.* (under press)

21. Goelzer, A., Bekkal Brikci, F., Martin-Verstraete, I., Noirot, P., Bessieres, P., Aymerich, S. and Fromion, V., (2008) *BMC Syst. Biol.* **2**:20, doi:10.1186/1752-0509-2-20

22. Kohn, K.W., (2001) *Chaos* **11**, 84-97

23. Kotoujansky, A., Lemaitre, M., and Boistard, P. (1982) *J. Bacteriol.* **150**, 122- 131

24. Miller, J. H. (1972) *Experiments in Molecular Genetics*, Cold Spring Harbor Laboratory Press, Cold Spring Harbor, New York

25. Moran, F., Nasuno, F., and Starr, P. (1968) *Arch. Biochem. Biophys.* **123**, 293-306

26. Reverchon, S., Rouanet, C., Expert, D., and Nasser, W. (2002) *J Bacteriol* **184**: 654-665.

27. Pfaffl, M.W. (2001) *Nucleic Acids Res.* **29**, e45

28. Funahashi A., Morohashi M., Kitano H., and Tanimura N. (2003) *Biosilico* **1**(5), 59-162

29. Britton, N. (2002) *Essential Mathematical Biology*, Springer, London

30. Tardy, F., Nasser, W., Robert-Baudouy, J., and Hugouvieux-Cotte-Pattat, N. (1997) *J. Bacteriol.* **179**, 2503-2511

31. Tagkopoulos, I., Liu ,Y.C., and Tavazoie, S. (2008) *Science* **320**, 1313-1317

32. Hugouvieux-Cotte-Pattat, N., Nasser, W., and Robert-Baudouy, J. (1994) *J. Bacteriol.* **176**, 2386-2392

33. Xiong, W., and Ferrell, J.E. (2003) *Nature* **426**, 460-465

34. Unpublished data.

**FIGURE LEGENDS**

Figure 1. Pectin degradation and catabolism in *Dickeya dadantii*. The upper part of the figure shows the detailed structure of pectin and the action of the different pectinases. The porins KdgM and KdgN, mediating the entry of unsaturated oligogalacturonate into the periplasm, and the transport systems TogMNAB and TogT, responsible for the entry of digalacturonate into the cytoplasm of the cells, are indicated. The secretion system of pectinases into the external medium (Out) is mentioned.

Figure 2. Simplified diagram of the regulatory network of Pel-KdgR coupled model of the bacterium *D. dadantii*. The action of KdgR on the *pel* genes are from data reported by Nasser et al. *(16)*.

Figure 3. Interaction graph representing the reactions pathway of the regulatory network of *pel*-KdgR coupled model of the bacterium *D. dadantii,* obtained using the CellDesigner tool.

Figure 4. Time course induction of Pel activity (blue squares) and *pelD* gene transcripts (circles) in *D. dadantii* strain 3937; the growth curve is indicated by black diamonds. Bacteria were grown at 30°C in liquid M63 minimal medium containing glucose (0.2% w/v) and PGA (0.01% w/v). Samples containing $10^9$ cells were taken every hour for RNA extraction and for Pel activity assays. Pel activity is expressed as μmoles of unsaturated product liberated per min and per ml of enzymatic extract. Each value represents the mean of two independent experiments. Bars indicate the standard deviation. Spline interpolation has been used to turn empirically measured transcript into a continuous function of time.
NB: *pelD* encodes one of the five major Pels (Pel A,B,C,D,E) of *D. dadantii*; similar results were also obtained with *pelB* and *pelE* genes.

Figure 5. The standard Michaelis Menten law fits the Pel enzymatic activity as a function of the PGA concentration. Assays were performed in minimal medium M63 buffer as described in the Experimental procedures section. Pel activity was normalized to the maximal Pel activity. The value of $K_m$ is classically extracted as the substrate concentration giving half of the maximum velocity. (Here $K_m = 1.2$ g/l). Each value represents the mean of three independent experiments. Bars indicate the standard deviation.

Figure 6. Time courses of the catabolism of PGA and the induction of Pel activity in *D. dadantii* strain 3937. Bacteria were grown at 30°C in liquid M63 minimal media containing glucose (0.2% w/v) and various quantities of PGA ((0.4 % w/v) in *(A)* and (0.2 % w/v) in *(B)*). 10 ml samples were taken every hour for PGA quantification and for Pel activity assays. At each time point, the remaining PGA concentration was expressed as the percentage of the initial PGA concentration. Experimental data are represented with squares (for Pel) and circles (for PGA) and are fitted with a continuous curve (solid line), computed from the maths model using parameters of Table 1. *A)* Initial amount of PGA (0.4% w/v) (4g/l) with $N_0 = 1.7x10^6$ bacteria inoculated corresponding to bacteria fractional volume $\rho_0 = 1.7x10^{-3}$. This data is used to determine the optimal set of unknown parameters. In this figure, the Pels synthesis for 2g/l amount of PGA is plotted with a green dashed line in order to observe the effect of changes in the initial PGA amount on Pels synthesis. *B)* Initial amount of PGA (0.2% w/v) (2g/l) with $N_0 = 7.3x10^7$ bacteria inoculated ($\rho_0 = 7.3x10^{-2}$). The maths model is used with the parameters determined to reproduce this experimental data. Each value represents the mean of two experiments. Bars indicate the standard deviation.
NB: Pel activity is expressed as μmoles of unsaturated product liberated/min/ml of enzymatic extract. Therefore, the numerical data for Pel activity is given by the expression: *Pel Act = $k_{cat}x$.*

Figure 7. The impact of added PGA, at the stationary phase of growth, on Pels production. Three flasks, containing the same volume of M63 minimal medium supplemented with glucose (0.2% w/v) and PGA (0.01 % w/v), were inoculated with a similar quantity of *D. dadantii* 3937 (*$N_0 = 4.6x10^8$*

bacteria inoculated, i.e. $\rho_0$=4.6x10$^{-1}$) and grown at 30°C for 24 h. Next, different quantities of PGA were added to the different growth media (red circles 0.01% w/v PGA added; blue diamonds 0.025% w/v PGA added; black squares 0.05% w/v PGA added). The cultures were then monitored until 50 h. 1 ml samples were taken every hour for Pel activity assays. Each value represents the mean of two independent experiments and bars indicate the standard deviation. The addition of PGA resulted in a slight resumption of bacterial growth and in a stimulation of Pel activity. Experimental data are fitted with a dashed line computed from the maths model using the parameters of Table 1. The dependence of the maximal level of Pels production on the amount of PGA added is shown in the inset of the figure.

Figure 8. Study of the sensitivity of Pel production dynamics for a 20% variability around parameters of Table 1. The blue line represents the best fit with no error on parameters. The green line (respectively the red line) represents the upper (respectively the lower) boundary for the set of parameters $p_{min}$ (respectively $p_{max}$) obtained with 10% decrease (respectively 10% increase) of parameters of Table 1 with the parameters of Table 2. The dashed green curve (respectively dashed red curve) represents the best fit obtained with the set of parameters $p_{max}$ (respectively $p_{min}$) and optimization of parameters of Table 2. The optimization results into a 50% increase (respectively 100% decrease) of $K_{d6}$, 10% increase (respectively 10% decrease) of $k_5$ and $\gamma$ and no variation on $\beta_1$ and $\varepsilon$. In dashed black, the KdgR dynamics is shown, using the parameters of Table 1. The temporary decrease of KdgR is due to the cytoplasmic dilution entailed by cell divisions during the exponential phase.

Figure 9. Variation of the occurrence time of some important events such as the maximal Pel production time ($t_{Pel\ max}$), the maximal KDG production time ($t_{KDG\ max}$) and the time of transition between the exponential and the stationary phases of bacteria growth ($t_{transition}$) as a function of the generation time of bacteria.

Figure 10. Prediction of the Pels dynamics in the *kdgK* mutant (blue line). One considers 0.4% w/v PGA and $N_0$ =1.7x10$^6$ inoculated bacteria ($\rho_0$=1.7x10$^{-3}$). For the wild type strain, $k_5$=50 $h^{-1}$. For the *kdgK* mutant $k_5$=2.5x$10^{-2}h^{-1}$ (The phosphorylation rate of KDG in the *kdgK* strain compared to the wild type strain has been divided by 2000). The figure also shows the time-evolution of KDG in the wild type, expressed in g/l (in dashed green).

Figure 11. The steady states of Pel activity are drawn in function of PGA (considered here as a parameter) by computing the stationary states of the system of eqs.(1)-(2) and eqs.(4)-(7), with $\rho=\rho_s$=10$^{-3}$. Stable states are shown in blue and unstable states are in red.

**Table 1:** List of parameter values deduced from independent experiments and literature sources.

| Param | Description | Value | units |
|-------|-------------|-------|-------|
| $\alpha_1$ | Degradation rate of Pel protein[1] | 4.94x10$^{-2}$ | $h^{-1}$ |
| $\alpha_2$ | Degradation rate of KdgR protein[1] | 3.47x10$^{-2}$ | $h^{-1}$ |
| $\beta_2$ | Maximum rate of synthesis of KdgR[2] | 7.00x10$^{-2}$ | $\mu M.h^{-1}$ |
| $K_m$ | Michaelis Menten constant for Pel.PGA complex[1] | 6.8 | $mM$ |
| $K_{d3}$ | Equilibrium dissociation constant for KdgR$_2$.KDG$_2$ complex[3] | 0.4 | $mM$ |
| $k_{cat}$ | Rate constant of the enzymatic Pel reactions in purified conditions[4] | 6x10$^4$ | $min^{-1}$ |

[1] The weight-characterized Michaelis Menten constant $K_M$=1.2 g/l has been measured in independent experiments as reported in this paper. From this value, one deduces the molar MM constant $K_m = K_M/P$, where P is the molecular weight of the galacturonate (P=176 Da Cf. reference *(30)*. For a theoretical discussion of the relation between $K_M$ and $K_m$, see also Supplemental Material, Section 3)

[2] Using the estimate that the number of KdgR$_2$ molecules measured during the growth phase is about 700 molecules per bacteria.

[3] Approximately 0.4 mM of KDG is required for a dissociation of 50% KdgR operator complexes. Cf. reference *(16)*. Let us note that the molecular weight of KDG (C$_6$O$_6$H$_{10}$) is 178 Da, which is close to the one of the galacturonic residue, 176 Da.

[4] The $k_{cat}$ of the 5 major Pel enzymes were measured in purified conditions, as reported in ref. *(30)*.

**Table 2:** List of parameters which are optimally determined by numerical simulations.

| Param | Description | Value | units |
|-------|-------------|-------|-------|
| $\varepsilon$ | Ratio of the rate constant of the enzymatic Pel reactions in the culture medium and the buffer[1] | 1 or 1/3 | - |
| $k_5$ | Phosphorylation rate of KDG | 150 | $h^{-1}$ |
| $\beta_1$ | Maximum rate of synthesis of Pels[2] | 30 | $\mu M\, h^{-1}$ |
| $\gamma$ | Cell membrane import rate | 60 | $h^{-1}$ |
| $K_{d6}$ | Equilibrium dissociation constant for KdgR$_2$.Pr-*pel* complex[3] | 4x10$^{-3}$ | $\mu M$ |

[1] In the simulations $\varepsilon\, k_{cat}$ represents the enzymatic reaction rate and $\varepsilon$ is let to vary as a free parameter bounded from above by *1*. The $k_{cat}$ of the 5 major Pel enzymes had been measured, as reported in ref. *(30)*, giving an average value of $6\times10^4\ min^{-1}$. As mentioned in the Discussion section, the value $\varepsilon$ =1 was found to be fairly good when a large set of polymer degradation reactions was considered (e.g. with maximum polymer length L=300), whereas $\varepsilon$ =1/3 was found to be optimal in the case L=2. The latter case corresponds to a simplified model where the enzymatic degradation occurs in a single reaction step.

[1] The obtained value of $\beta_1$ corresponds to a maximal synthesis rate of 5 proteins per second and per cell (with $v_{cell}$ = 1 ($\mu$m)$^3$). So if mainly 5 *pel* genes are involved in this production, the maximal production rate corresponds to 1 protein/s/cell.

[2] In ref. *(16)*, values for the dissociation constant for KdgR-operator of *pel* genes were measured and found typically between 10$^{-3}$ and 10$^{-4}$ $\mu M$.

Figure 1

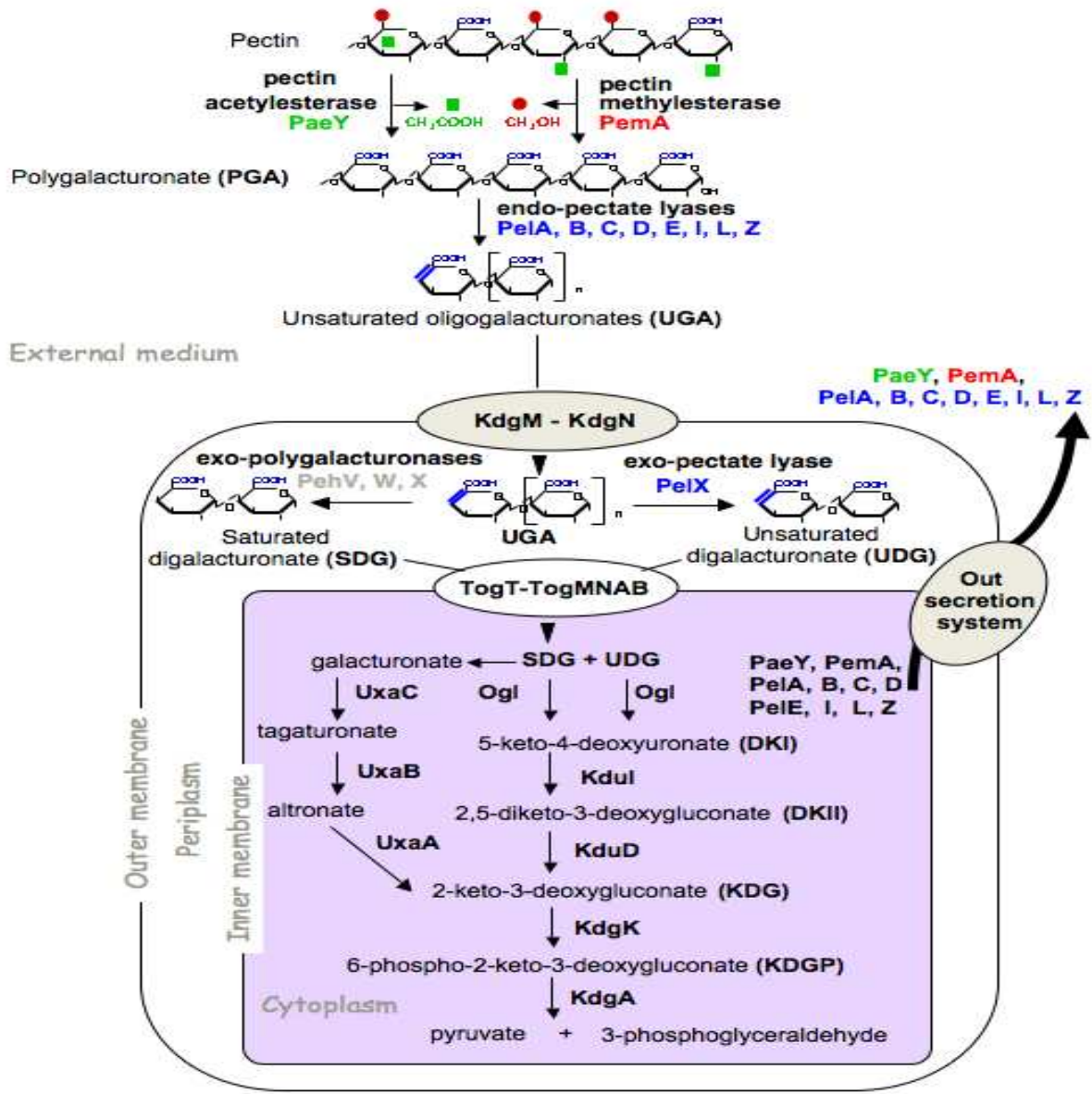

Figure 2

Figure 3

Figure 4

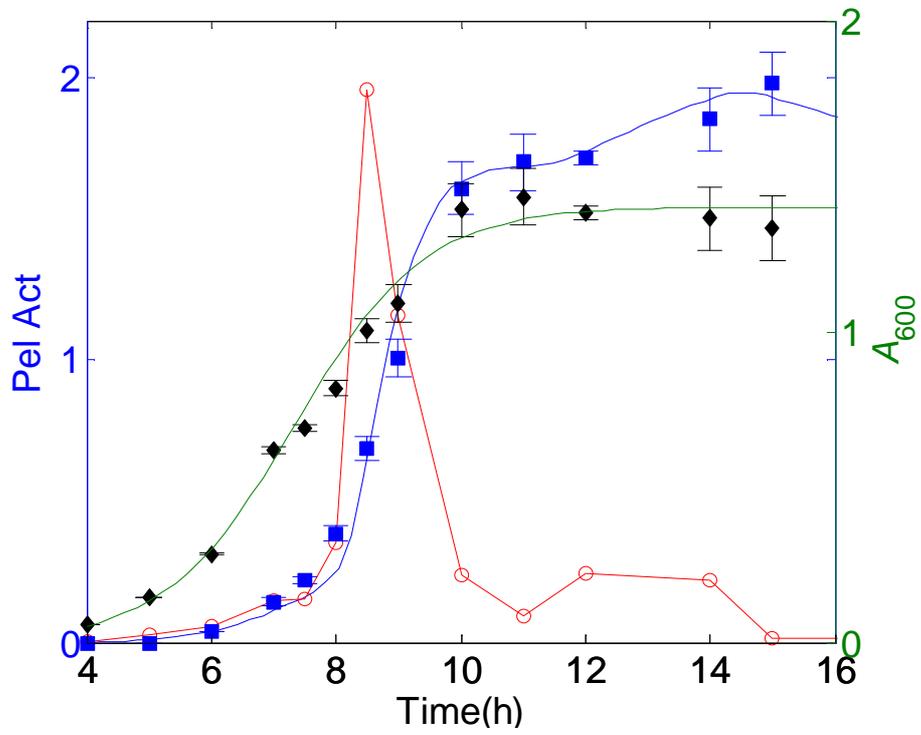

Figure 5

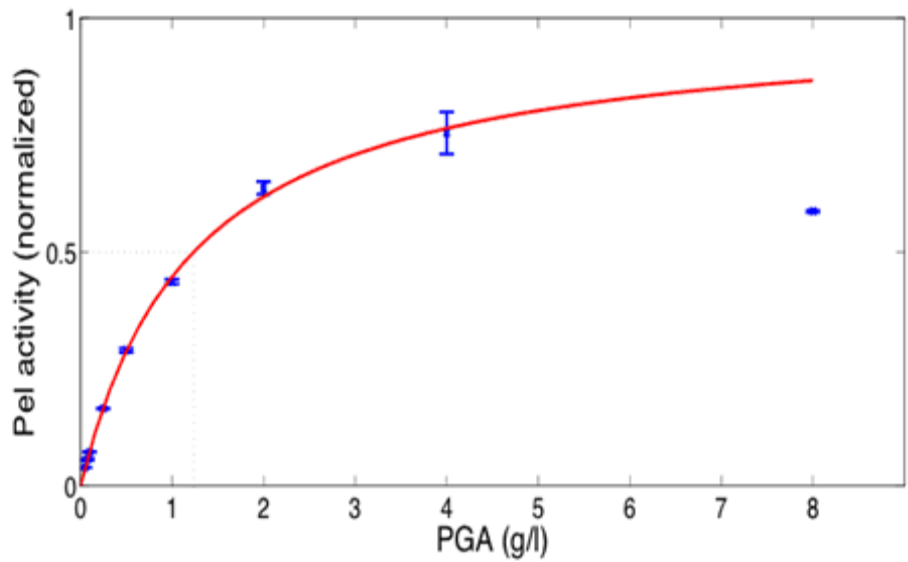

Figure 6

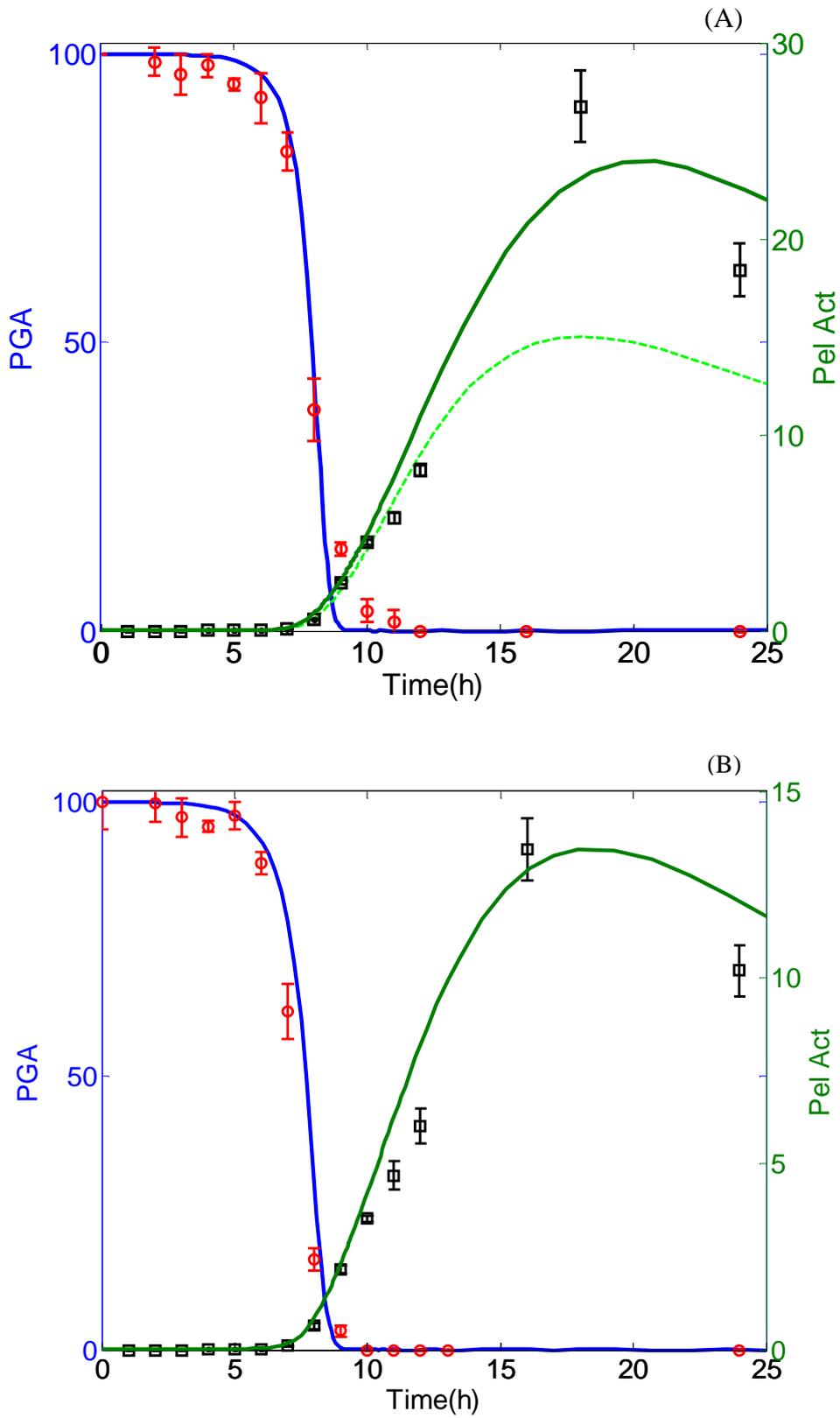

Figure 7

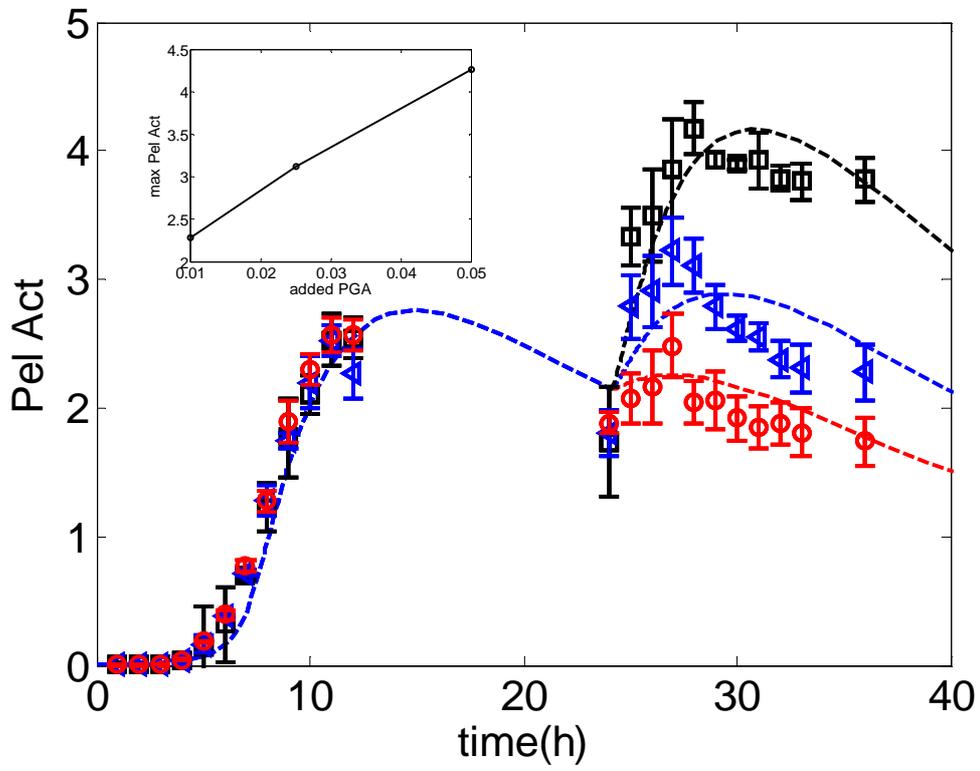

Figure 8

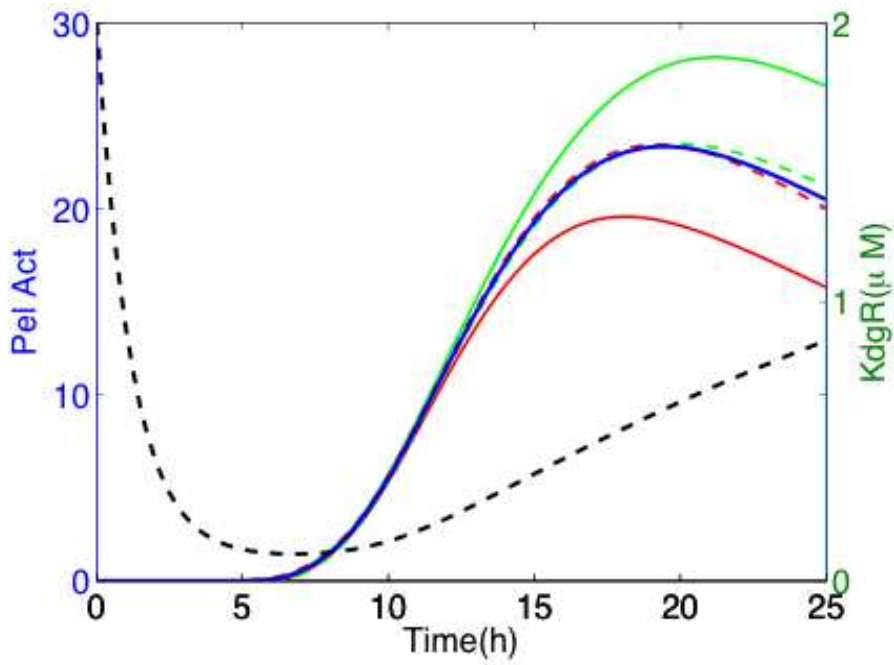

Figure 9

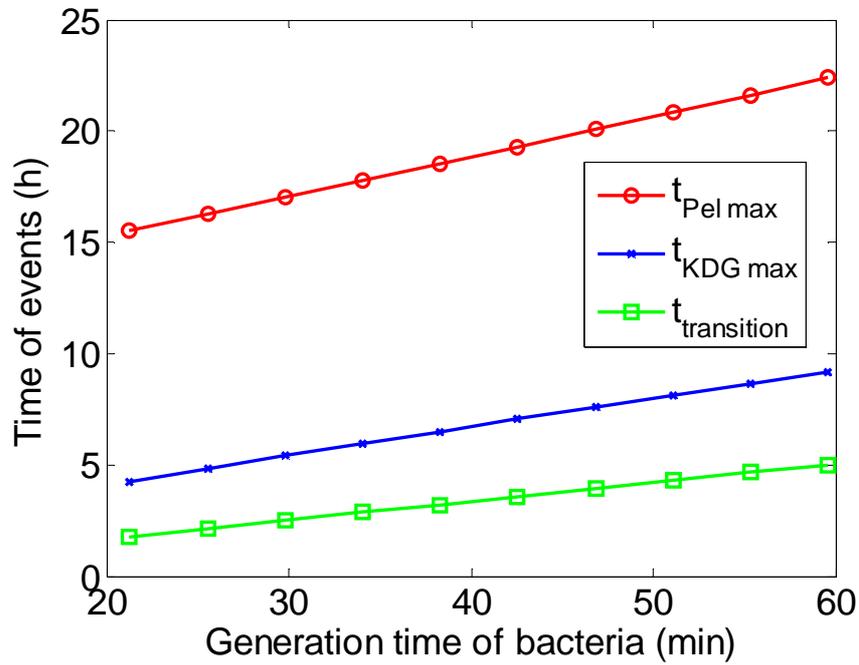

Figure 10

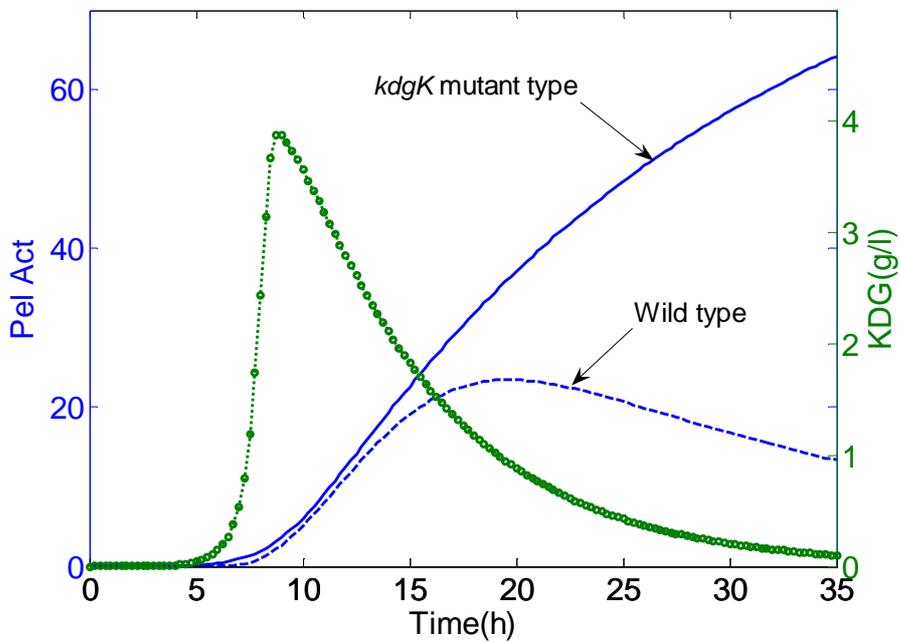

Figure 11

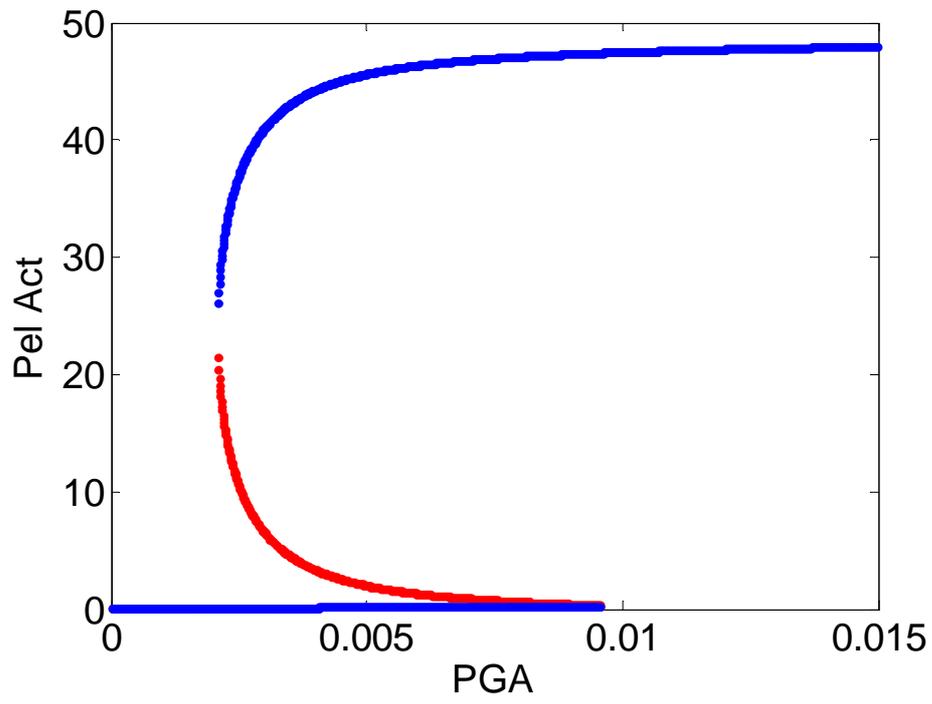